\documentclass[pra,twocolumn,preprintnumbers,amsmath,amssymb,floatfix,
superscriptaddress,showpacs]{revtex4-1}
\usepackage{graphicx}
\usepackage{graphics}
\usepackage{psfrag}
\usepackage{hyperref}
\usepackage{multirow}
\usepackage[sort&compress]{natbib}
\usepackage{color}
\usepackage{slashed}

\newcommand{\vare}{\varepsilon}

\newcommand{\eq}[1]{Eq.~(\ref{#1})}
\newcommand{\fig}[1]{Fig.~\ref{#1}}
\newcommand{\tab}[1]{Tab.~\ref{#1}}
\newcommand{\imag}{\text{i}}
\renewcommand{\sec}[1]{Sec.~\ref{#1}}

\begin{document}

\hypersetup{pdftitle={Sarma phase in relativistic and non-relativistic systems}}
\title{Sarma phase in relativistic and non-relativistic systems}
\date{\today}
\author{I. Boettcher}
\affiliation{Institute for Theoretical Physics, Heidelberg University, D-69120 Heidelberg, Germany}
\author{T. K. Herbst}
\affiliation{Institute for Theoretical Physics, Heidelberg University, D-69120 Heidelberg, Germany}
\author{J. M. Pawlowski}
\affiliation{Institute for Theoretical Physics, Heidelberg University, D-69120 Heidelberg, Germany}
\affiliation{ExtreMe Matter Institute EMMI, GSI Helmholtzzentrum f\"{u}r Schwerionenforschung mbH, D-64291 Darmstadt, Germany}
\author{N. Strodthoff}
\affiliation{Institute for Theoretical Physics, Heidelberg University, D-69120 Heidelberg, Germany}
\author{L. von Smekal}
\affiliation{Theoriezentrum, Institut f\"{u}r Kernphysik, TU Darmstadt, D-64289
Darmstadt, Germany}
\affiliation{Institut f\"{u}r Theoretische Physik, Justus-Liebig-Universit\"{a}t, D-35392 Giessen, Germany}
\author{C. Wetterich}
\affiliation{Institute for Theoretical Physics, Heidelberg University, D-69120
Heidelberg, Germany}
\affiliation{ExtreMe Matter Institute EMMI, GSI Helmholtzzentrum f\"{u}r
Schwerionenforschung mbH, D-64291 Darmstadt, Germany}

\begin{abstract}
We investigate the stability of the Sarma phase in two-component fermion systems
in three spatial dimensions. For this purpose we compare strongly-correlated
systems with either relativistic or non-relativistic dispersion relation: 
relativistic quarks and mesons at finite isospin density and spin-imbalanced
ultracold Fermi gases. Using a Functional
Renormalization Group approach, we resolve fluctuation effects onto the
corresponding phase diagrams beyond the mean-field approximation. We find that
fluctuations induce a second-order phase transition at zero temperature, and
thus a Sarma phase, in the relativistic setup for large isospin chemical
potential. This motivates the investigation of the cold atoms setup with
comparable mean-field phase structure, where the Sarma phase could then be
realized in experiment. However, for the non-relativistic system we find the
stability region of the Sarma phase to be smaller than the one predicted from
mean-field theory. It is limited to the BEC side of the phase diagram, and the
unitary Fermi gas does not support a Sarma phase at zero temperature. Finally,
we propose an ultracold quantum gas with four fermion species that has a good
chance to realize a zero-temperature Sarma phase. 
\end{abstract}

\pacs{05.10.Cc, 11.10.Hi, 67.85.Lm}

\maketitle

\section{Introduction}

Understanding the pairing mechanisms in fermionic many-body systems is a key
step towards bridging the gap between microscopic models and macroscopic
phenomena. A particularly interesting question concerns the stability of
superfluidity in the presence of mismatching Fermi surfaces. Such an asymmetry
between the pairing partners is realized in electronic materials in an external
magnetic field \cite{PhysRevLett.9.266, Chandrasekhar, Sarma19631029,
PhysRev.135.A550, Larkin:1964zz}, and is expected to be found in neutron stars 
\cite{Lombardo:2000ec, RevModPhys.75.607, Alford:2007xm, Page:2013hxa,
Gezerlis:2014efa, Kruger:2014caa}. With ultracold atoms this situation can
easily be simulated by introducing a population imbalance between different
hyperfine states. In a relativistic, QCD-related setting, non-vanishing isospin
chemical potential similarly introduces an imbalance between different quark
flavors, up and down. Alternatively, we may consider the relativistic isospin
chemical potential as maintaining a balance between up and anti-down quarks with
pairing in a superfluid pion condensate. The pair-breaking population imbalance
is then introduced by the symmetric quark or baryon chemical potential.

While it is not possible to study, for example, the pairing mechanisms in
neutron stars in table-top experiments, the high experimental control and
accessibility of ultracold quantum gases makes them ideal setups to shed new
light on superfluidity and its breakdown \cite{ketterle-review, gurarie-review,
RevModPhys.80.885, RevModPhys.80.1215, Zwerger-book}. In particular,
preparations of the spin-imbalanced BCS-BEC crossover allow the tuning of system
parameters almost at will \cite{Partridge27012006, Zwierlein27012006,
PhysRevLett.97.030401, PhysRevLett.97.190407,Salomon,
Navon07052010,Chevy:2010zz}. It is then interesting to study whether there is a
parameter regime of the non-relativistic model that corresponds to a system
relevant for nuclear or possible quark matter at high densities.
Apart from the physical similarities of these systems, a positive answer to
this question is expected based on the observation that the mean-field phase
diagrams in both cases look strikingly similar. 
To reach a conclusive statement, however, fluctuations beyond the
mean-field approximation need to be taken into account, which is the aim of the
present work.

In connection with the breakdown of superfluidity, the possible existence of a
so-called Sarma phase \cite{Sarma19631029} has gained a lot of interest
recently \cite{PhysRevLett.90.047002, PhysRevA.67.053603, Shovkovy2003205,
PhysRevLett.91.247002, PhysRevA.70.033603, PhysRevLett.95.060401,
PhysRevA.74.033606, Gubbels:2006zz, PhysRevLett.96.060401, PhysRevD.74.036005,
2006PhLB..637..367K, 2007NatPh...3..124P, PhysRevA.75.033608,
PhysRevLett.98.160402, PhysRevLett.100.140407, PhysRevB.79.024511,
PhysRevB.79.024511, 0034-4885-73-7-076501, Gubbels:2013mda}.
The Sarma phase is a homogeneous superfluid phase with gapless fermionic
excitations. To understand its origin we consider a gas of two species of
fermions, labelled by an effective ``spin'' 1 and 2, with a chemical
potential imbalance $\delta\mu=(\mu_1-\mu_2)/2\geq0$  between them.
After including renormalization effects on the propagator of fermionic
quasiparticles, we can infer their dispersion relation from the quadratic part
of the spin-imbalanced effective Lagrangian. It typically splits into two lowest
branches given by
\begin{align}
 \label{eq:Int1} E_{p}^{(\pm)} &= \sqrt{\vare_p^2+\Delta^2} \pm\delta\mu,
\end{align}
where $\vare_p$ is the microscopic dispersion relation of particles, and 
$\Delta$ is the pairing gap. 

\begin{figure}
  \includegraphics[height=.191\textheight]{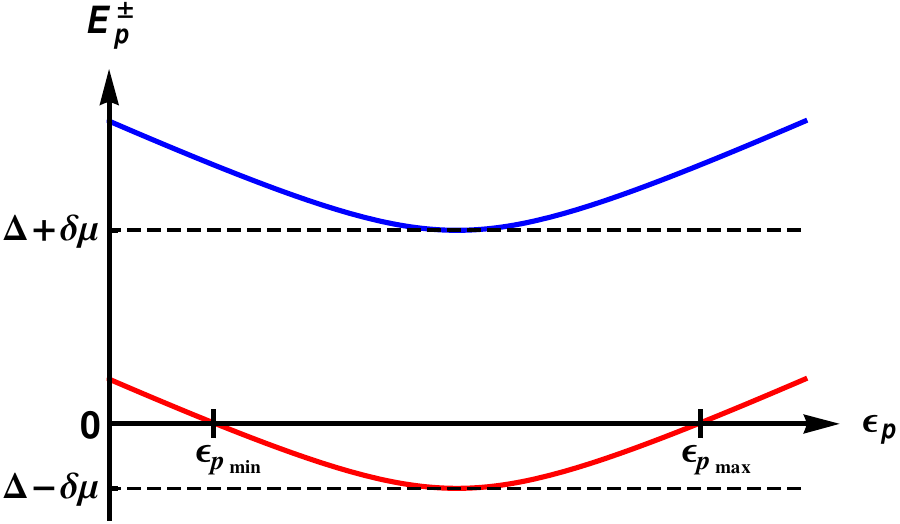}
  \caption{The two lowest branches of the dispersion relation, \eq{eq:Int1},
    relevant for the Sarma transition. Increasing the imbalance
    $\delta\mu$, the lowest  branch extends below zero, yielding gapless
    excitations around the Fermi surfaces at the corresponding momenta $p_{\rm
    min}$ and $p_{\rm max}$.
    Note that for $\min_p \vare_p >0$ the minimal $\vare_{p_{\rm min}}$ can
    become negative, and the Sarma phase appears with only one Fermi surface
    in this case.}
  \label{fig:SarmaDispersion}
\end{figure}

A Sarma phase is characterized by a non-vanishing gap $\Delta$ and
the parameters in \eq{eq:Int1} are tuned such that the
lower branch becomes negative in a momentum interval $p_{\rm min} < p < p_{\rm
max}$\,, see \fig{fig:SarmaDispersion}. Accordingly, this interval
becomes occupied even at zero temperature, and we find gapless excitations
around the built-up Fermi surfaces at $p_{\rm min}$ and $p_{\rm max}$. For the
remaining momenta, fermionic excitations are gapped.
For non-zero temperature the Fermi surfaces are smeared out and a sharp
distinction between the unpolarized superfluid and the Sarma phase is not
possible.  Hence, we speak of a Sarma crossover in this case. The Sarma phase, or special
cases of it, is also referred to as interior gap superfluid, breached pair
phase, or magnetized superfluid in the literature.

\begin{figure}
    \includegraphics[height=.2\textheight]{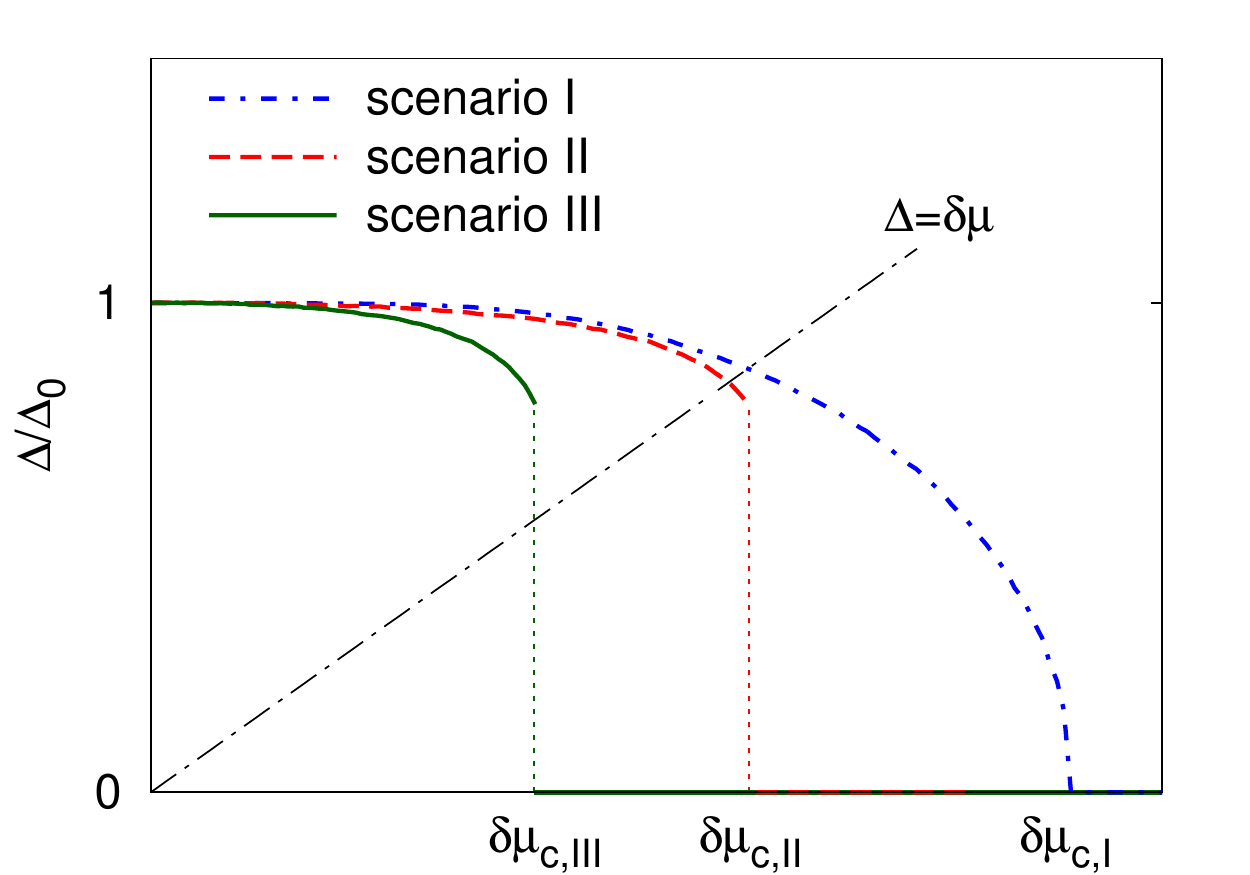}  
    \caption{The three possible scenarios for the Sarma condition
    $\Delta=\delta\mu$; see discussion below \eq{eq:Int2} for details.
    In the case of a second-order superfluid phase transition the criterion is
    always fulfilled for some $\delta\mu$ (Scenario I), whereas the
    size of the critical gap at a first-order transition decides whether it is
    fulfilled (Scenario II) or not (Scenario III).}
  \label{fig:scenarios}
\end{figure}

The criterion for a zero crossing of the lower branch in \eq{eq:Int1}, and thus
for the onset of the Sarma phase, is equivalent to
\begin{align}
 \label{eq:Int2} \delta\mu > \min_p \sqrt{\vare_p^2+\Delta^2}\,.
\end{align}
We emphasize again that this equation is understood in terms of renormalized
single-particle quantities. Assuming for simplicity that $\min_p \vare_p=0$, we
then arrive at the condition $\delta\mu>\Delta$.
Then there are three possible scenarios for a spin-imbalanced system with $\Delta>0$,
which decide over the fate of the Sarma phase.
By increasing $\delta\mu$, we make pairing less favorable, and superfluidity
generically breaks down at a critical imbalance $\delta\mu_{\rm c}$.
If this happens continuously, i.e. by means of a second-order
phase transition, the Sarma criterion is necessarily fulfilled 
somewhere, since $\Delta \to 0$ (scenario I). This is
depicted by the blue, dot-dashed line in \fig{fig:scenarios}.
At a first-order phase transition, on the other hand, the gap jumps 
from a critical value $\Delta_{\rm c}>0$ to zero. For $\delta\mu_{\rm c}
>\Delta_{\rm c}$ a Sarma phase exists (scenario II; red, dashed line in
\fig{fig:scenarios}), whereas the required condition cannot be fulfilled for
$\delta\mu_{\rm c}<\Delta_{\rm c}$ (scenario III; green, solid line). We see
that the existence of a Sarma phase at a second-order transition line is a
universal feature, whereas it becomes non-universal in the vicinity of a 
first-order transition line.

In experiments with ultracold atoms the Sarma phase can be inferred from a
non-monotonous or non-continuous momentum distribution after time-of-flight
expansion \cite{Gubbels:2013mda}. At non-zero temperature, the sharp features in
the momentum distribution are smeared out. The Sarma phase also shows up in
shell-structured in-situ density images, where the polarized superfluid
manifests itself in a population imbalance between the spin species
\cite{Gubbels:2006zz}: If the transition to the normal gas is of first order, an
intermediate population imbalanced superfluid region in the cloud which smoothly connects
to the balanced superfluid, indicates the Sarma phase. If the transition is of
second order, the superfluidity of the population imbalanced region can be
probed by the excitation of vortices. The presence of Fermi surfaces is also
expected to induce metallic features in the superfluid, which are observable
in its transport properties. This makes the system an unconventional superfluid.
The transport properties of neutron stars are known to be closely linked to
their constitution and life time. A possible Sarma phase is thus of relevance
for interpreting the stellar evolution.

As discussed above, the two systems studied in the following look
very similar at first glance and indeed it is found that their mean-field phase
diagrams agree qualitatively. Upon closer inspection, however, especially the
bosonic sectors of the two theories differ. Fluctuation contributions from
this sector are not accounted for in a mean-field approximation, but may lead to
strong modifications of the phase diagram. Notably, the relativistic system
shows a richer phase structure, including a Sarma phase at low
temperature once fluctuations are taken into account, see
\sec{sec:Relativistic}. One can then ask whether this is also true in the
non-relativistic setting, where such additional phases are potentially
accessible in experiment.

To study spin-imbalanced systems beyond the mean-field approximation, we employ
the Functional Renormalization Group (FRG), which enables the systematic
inclusion of fluctuations.
In particular, it naturally incorporates the feedback of order
parameter fluctuations onto the full effective potential. As a consequence,
physical observables show the correct beyond mean-field scaling at second-order
phase transitions. 
Additionally, the FRG is free of the sign problem, which hampers
Quantum Monte Carlo studies of spin-imbalanced systems \cite{Braun:2012ww}.
Hence, the full phase diagrams of both the spin-imbalanced non-relativistic and
the isospin-asymmetric relativistic system are accessible. For extensive reviews
on the method see Refs. \cite{Berges:2000ew, Gies:2006wv, Schaefer:2006sr,
Pawlowski20072831, Delamotte:2007pf, Kopietz2010, Metzner:2011cw,
Braun:2011pp}.
Comprehensive introductions to the FRG approach for QCD-like models and
the BCS-BEC crossover, respectively, can be found in Refs.
\cite{vonSmekal2012179, Schaefer:2011pn} and
\cite{Scherer:2010sv,Boettcher201263}.

To highlight the impact of bosonic fluctuations, we
compare results in the mean-field approximation to those obtained with the FRG.
In many cases, mean-field theory predicts a first-order breakdown of
superfluidity due to spin-imbalance at $T=0$. Including fluctuations, this
first-order transition can turn into a continuous one. This interesting effect
has indeed been found in FRG studies of two-dimensional Hertz--Millis type
actions \cite{PhysRevLett.103.220602}, and a non-relativistic spin-imbalanced
Fermi gas on the BCS-side of the crossover in two spatial dimensions
\cite{PhysRevX.4.021012}. In the present analysis we find such a smoothing of
the transition in the relativistic model, whereas it is absent in the
non-relativistic setting.

This paper is organized as follows. In \sec{sec:Relativistic} we consider
the relativistic system, where bosonic fluctuations induce a Sarma phase
close to the breakdown of charged pion condensation at zero temperature. We then
consider the non-relativistic analog in \sec{sec:NonRelativistic}.
After discussing our approximation we investigate the stability of the Sarma
phase in the unitary Fermi gas at zero and finite temperature, and then turn to
the imbalanced BCS-BEC crossover at zero temperature. We draw our conclusions in
\sec{sec:Concl}.

\begin{figure*}
  \includegraphics[width=.45\textwidth]{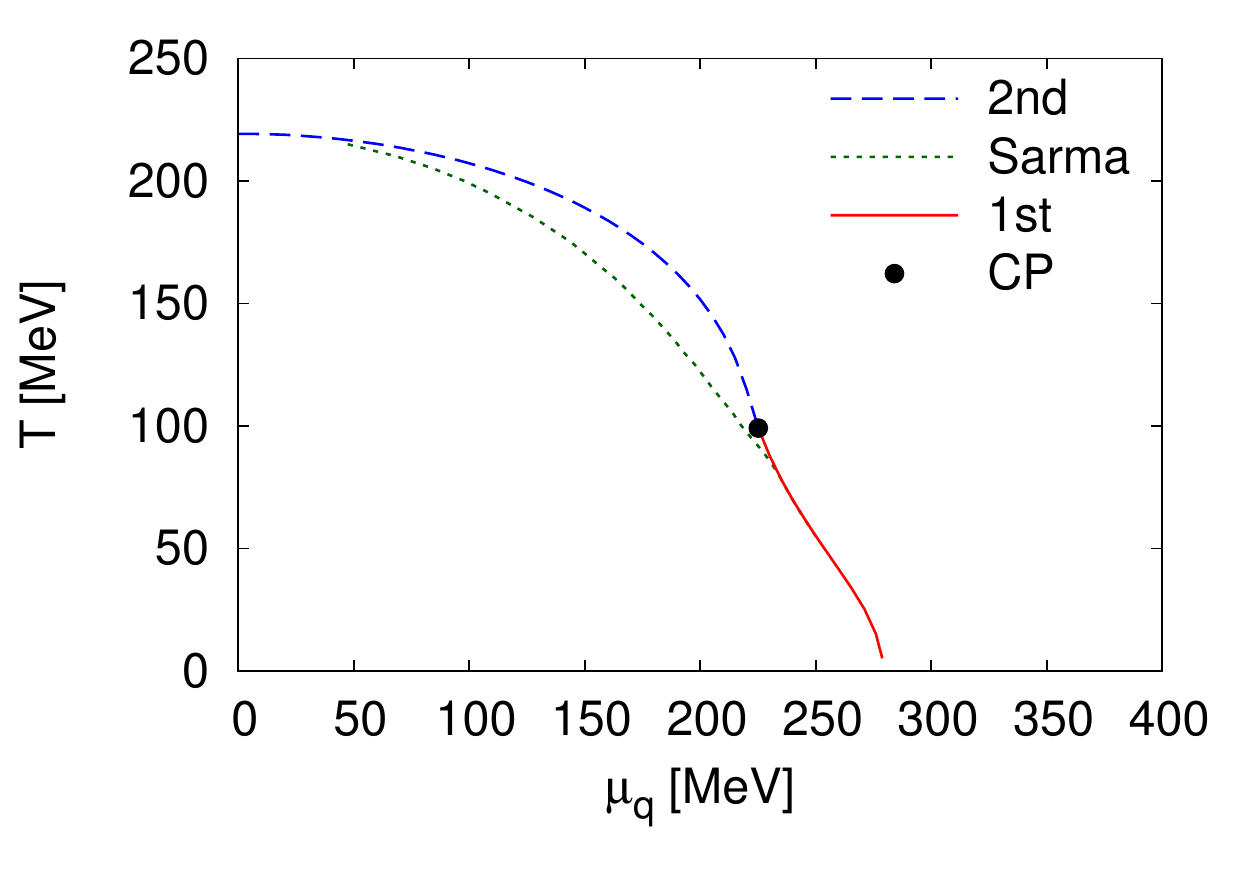}
  \includegraphics[width=.45\textwidth]{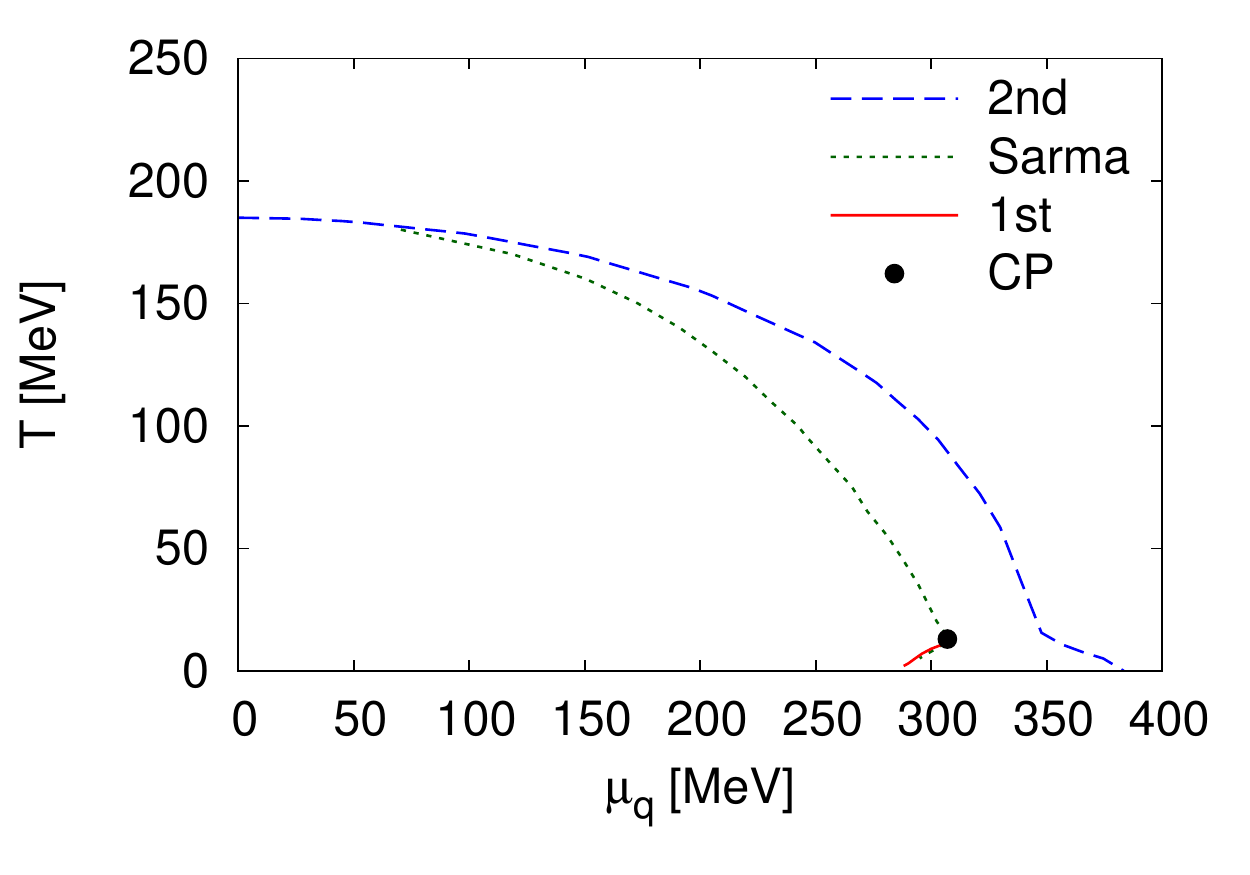}
  \caption{Phase structure of the quark-meson model for $\mu_I=m_\pi$  in
    mean-field approximation (MFA, left) and from the FRG (right). The
    Sarma phase occurs in the region between the dotted and dashed lines. While
    the MFA result looks very similar to the one of the UFG, cf.
    \fig{fig:UFG_nonrel}, the FRG result features a Sarma phase at $T=0$. }
  \label{fig:sarmarel}
\end{figure*}

\section{Relativistic System} 
\label{sec:Relativistic}

In this section we investigate the fate of the Sarma phase in a relativistic 
system. To this end, we employ a quark-meson model, which is frequently used as
a low-energy effective model for QCD, with quarks and mesons as effective
degrees of freedom \cite{Ellwanger:1994wy, Jungnickel:1995fp, Berges:1997eu,
Schaefer:2004en,Herbst:2010rf, PhysRevD.88.014007}. 
Here we introduce both a finite quark, $\mu_q$, and isospin, $\mu_I$, chemical
potential. The quark chemical potential, defined as one third of the baryon
chemical potential, induces an imbalance between quarks and anti-quarks. In
contrast, the isospin chemical potential induces an imbalance between the quark
flavors, up and down.  Similar to the non-relativistic case discussed in
\sec{sec:NonRelativistic}, this setup allows to study the impact of mismatched
Fermi spheres on fermion pairing. In fact, at low temperature and finite
densities, the system describes a superfluid in the BCS-BEC crossover
\cite{PhysRevLett.86.592}, similar to the non-relativistic case discussed below:
At moderately large $|\mu_I|>m_\pi/2$ charged pions condense and form a Bose
condensate.  The ground state then is a superfluid of pions. In the
limit of large  $|\mu_I|$, on the other hand, Cooper-pairing between  quarks
and antiquarks sets in.
Also in this case the relevant channel carries the quantum numbers of a pion.
Hence we expect a smooth crossover from BEC- to BCS-like pairing as $\mu_I$ is
increased.

Moreover, the case of a pure isospin chemical potential, i.e. vanishing quark
chemical potential, is one example for a QCD-like theory without a fermion sign
problem.
The latter represents the main obstacle for studying the phase diagram of QCD at
finite quark chemical potential using Lattice Monte Carlo methods 
\cite{deForcrand:2010ys}.
However, the situation of both, a finite quark and finite isospin chemical
potential, is also of direct physical interest in the context of heavy ion
collisions or quark matter inside neutron stars.

The omission of a possible diquark condensate and the absence of baryonic
degrees of freedom constitute natural limitations of the capability of this
model to describe QCD at high densities. Here, however, we are mainly interested
in the similarities of this relativistic model with its non-relativistic
counterpart discussed below. Hence, such QCD-related limitations are of no
concern for the present work.

The model is described by a Lagrangian of the form \cite{Kamikado:2012bt}
\begin{equation}
\label{eq:Lagrangianqmmuf}
\begin{split}
 {\cal L}=&\  \bar\psi\left(\slashed \partial +g(\sigma+\imag\gamma^5
\vec \pi\vec\tau)-\gamma_0\mu_q-\gamma_0\tau_3 \mu_I\right)\psi\\
  &+\tfrac{1}{2}(\partial_{\nu} \sigma)^2  +\tfrac{1}{2}(\partial_{\nu}  
  \pi_0)^2 + U(\chi,\rho)- c \sigma \\ 
  &+ \tfrac{1}{2} (\partial_\nu + 2
    \mu_I\delta^0_\nu)\pi_+(\partial_\nu - 2\mu_I
    \delta^0_\nu)\pi_-\,,  
\end{split}
\end{equation}
where $\tau_i$ denote Pauli matrices in flavor space and we define
invariants $\chi\equiv \sigma^2 + \pi_0^2$ (with $\pi_0\equiv \pi^3$),
$\rho\equiv \pi_+ \pi_-$ (with $\pi_\pm=\pi_1\pm\imag \pi_2$) and $\psi =
(\psi_u, \psi_d)^T$. The four (real) bosonic degrees of freedom are given by the
isospin-singlet ($\sigma$) and -triplet ($\vec \pi$), which combine to a
four-component $(2,2)$ representation of the chiral $SU(2)_L~\times~SU(2)_R$
symmetry of the theory. 

At finite isospin chemical potential the effective potential in general is a
function of both invariants $\chi$ and $\rho$. Its minimum $(\chi_0,\,\rho_0)$
determines the phase structure of the system, where a finite value of
$\chi_0=\langle \chi\rangle$ signals broken chiral symmetry and a finite value
of $\rho_0=\langle \rho\rangle$ signals a phase of charged pion condensation.
This value can then be used to define the gap parameter as $\Delta^2
\equiv g^2\rho_0$.
The following analysis focuses on the charged pion condensation phase and its
disappearance with increasing quark chemical potential.

Within the framework of the FRG we investigate the model \eq{eq:Lagrangianqmmuf}
in the local potential approximation (LPA), where only a
scale-dependent effective potential is considered. However, the full field
dependence of the effective potential is taken into account by expanding it on a
two-dimensional grid in field space \cite{Strodthoff:2011tz}. For a
comprehensive description of the phase structure of this model as a function of
the three external parameters  $(T,\, \mu_q,\, \mu_I)$ as obtained with the FRG,
as well as for a more detailed description of the truncation and implementation,
we refer the reader to \cite{Kamikado:2012bt}. Here we only briefly recapitulate
the features most relevant for the present investigation. 

For sufficiently large isospin chemical potential and sufficiently small quark
chemical potential there is a phase of charged pion condensation. At zero
temperature the phase diagram is strongly constrained by the Silver Blaze
property \cite{Cohen:2004qp, Kamikado:2012bt}, which prohibits a dependence of
the partition function on the chemical potential until the latter exceeds the
mass of the lowest excitation it couples to. At vanishing quark chemical
potential for example, this entails that the onset of pion condensation is found
at $\mu_I = m_\pi/2$. At fixed $\mu_I>m_\pi/2$ with increasing quark chemical
potential $\mu_q$ the charged pion condensation phase finally breaks down. Interestingly,
for $\mu_I>0.79 m_\pi$ the full calculation including bosonic fluctuations shows an 
additional first-order transition at small and vanishing temperatures inside 
the pion condensation phase close to its phase boundary \cite{Kamikado:2012bt}.

One possible interpretation for this transition is a first-order transition to
a Sarma phase, corresponding to scenario II in \fig{fig:scenarios}, and hence
the existence of  a stable Sarma phase at vanishing temperature. As outlined
above, the definition of the Sarma phase relies on the notion of quasiparticle
dispersion relations, which, for the Lagrangian given in \eq{eq:Lagrangianqmmuf}, take the
form
\begin{align}
 \vare_p = \sqrt{\vec p^2+m_\psi^2}-\mu_I\,,
\end{align}
where $m_\psi^2=g^2\chi$. For $\mu_I >m_\psi$ we have $\min_p\vare_p=0$ 
\cite{Kamikado:2012bt,Ebert:2005cs}, and the criterion for the stability of the
Sarma phase, \eq{eq:Int2}, reduces to $\Delta < \mu_q$. In particular,
this is true in the case of restored chiral symmetry. To investigate the 
appearance of a Sarma phase in more detail, we study slices of the
three-dimensional phase diagram at a fixed value of $\mu_I=m_\pi$. As remarked above,
for this value a first-order transition inside the pion condensation occurs 
upon increasing $\mu_q$. Estimating the location of the BCS-BEC crossover via
the simple criterion of the zero-crossing of $\min_p \vare_p$, i.e.
$\mu_I=m_\psi$, translates into a value of $0.82 m_\pi$. Hence the choice
$\mu_I=m_\pi$ corresponds to a point just on the BCS-side of the crossover. 

The corresponding $(\mu_I-T)$-phase diagrams are shown in \fig{fig:sarmarel}, 
where the left panel shows the outcome of a mean-field calculation which can be
obtained in a consistent way from the FRG by including only fermionic
contributions to the flow, cf. \sec{sec:Trunc} below. For small temperatures the
boundary of the pion condensation phase is a first-order transition line. This
is the analog of the Chandrasekhar--Clogston transition
\cite{PhysRevLett.9.266, Chandrasekhar} in non-relativistic Fermi gases. It
turns second order for larger temperatures in a  multicritical point which
could become a Lifshitz point if  inhomogeneous Fulde--Ferrell-Larkin--Ovchinnikov phases occur
\cite{PhysRev.135.A550, Larkin:1964zz, Chevy:2010zz,
Roscher:2013cma, Braun:2014ewa}. This situation is analogous to the
inhomogeneous phases discussed for chiral symmetry restoration at finite baryon
chemical potential in QCD, for a recent review see \cite{Buballa:2014tba}.
As outlined in the introduction, the second-order transition line at larger
temperatures is accompanied by a stable Sarma phase. The Sarma phase, however,
does not extend to the zero temperature axis, because the Chandrasekhar--Clogston
limit is  reached before a possible Sarma transition could occur at low
temperatures in the mean-field calculation.

In contrast, the right panel of \fig{fig:sarmarel} shows the full result
including bosonic fluctuations in LPA. Here the phase boundary of the pion
condensation phase remains second order throughout the whole phase diagram.
However, an additional first-order transition arises inside the pion
condensation phase. As the condensate jumps to a sufficiently low value across
the phase boundary, the Sarma criterion is satisfied. Therefore, unlike in the
mean-field calculation, the Sarma  phase now extends down to zero temperature in
the calculation including bosonic fluctuations.
As discussed above, mean-field studies of chiral systems, however, suggest that
the phase structure at low temperature is altered once inhomogeneous phases are taken into
account. This effect may persist when fluctuations are included, but for the system under
consideration no results are available thus far. In fact, the study of inhomogeneous phases beyond
MFA poses a sophisticated task, see e.g. \cite{PhysRevB.80.014436, Braun:2014fga} for recent
developments within the FRG. For instance, within a derivative expansion the vanishing of the
bosonic wave function renormalization may signal the onset of inhomogeneous condensation
\cite{PhysRevB.80.014436}.

Furthermore, on the mean-field level, the phase diagrams of the
relativistic system, \fig{fig:sarmarel} (left), and the unitary Fermi gas (UFG),
\fig{fig:UFG_nonrel} (upper lines), look strikingly alike. In fact, the phase
structure of the imbalanced Unitary Fermi Gas has become experimentally
accessible by now. The existence of a similar Sarma phase at low $T$ in the
non-relativistic system could hence be checked experimentally. This serves as
our motivation to include fluctuations in the non-relativistic system and study
its phase structure in \sec{sec:NonRelativistic} below.

\section{Non-relativistic System} 
\label{sec:NonRelativistic}

As the non-relativistic realization of the system under consideration, we study
a system of ultracold two-component fermions close to a broad Feshbach resonance
(FR). Its description in terms of the two-channel model is built on a Grassmann
field $\psi_\sigma$, one complex bosonic field $\phi$ and
the microscopic Lagrangian \cite{gurarie-review,Diehl:2005ae,Zwerger-book}
\begin{align}\label{eq:Lnonrel}
 \nonumber \mathcal{L} &=  \sum_{\sigma=1,2} \psi^*_\sigma  \Bigl(\partial_\tau 
- \frac{\nabla^2}{2M_\sigma}-\mu_\sigma\Bigr)\psi_\sigma -
g\Bigl(\phi^*\psi_1\psi_2+\text{h.c.}\Bigr)\\
 &\mbox{ } +  \phi^*\Bigl(Z_\phi \partial_\tau - A_\phi 
\frac{\nabla^2}{4M}\Bigr)\phi +\nu_\Lambda\phi^*\phi\,.
\end{align} 
The two species of fermions couple to chemical potentials $\mu_\sigma$, which
in general are different. We assume the 1-atoms to be the majority species and
write
\begin{align}
 \mu_1 = \mu+\delta\mu,\ \mu_2=\mu-\delta\mu\,,
\end{align}
with spin-imbalance $\delta\mu=h=(\mu_1-\mu_2)/2\geq0$. The quantity $h$ is also
referred to as Zeeman field. We assume mass balance in the following and
choose units such that $\hbar=k_{\rm B}=2M_\sigma=2M=1$ for the non-relativistic
analysis. 

The parameter $\nu_\Lambda\propto (B-B_0)$ is related to the detuning from the
FR, and eventually allows the computation of the s-wave scattering length, $a$,
of the system. The Feshbach coupling $g^2\propto\Delta B$ corresponds to the
width of the FR.
We assume the FR to be broad in the following, such that the two-channel model
in \eq{eq:Lnonrel} is equivalent to a single-channel model of fermions.

The self-interaction of the bosonic degree of freedom is encoded in the
effective potential, $U(\rho=\phi^*\phi)$. At the microscopic scale we have
$U(\rho)=\nu_\Lambda\rho$, but the $\rho$-dependence is changed  substantially
when including fluctuations. In the following, we compute the effective
potential $U(\rho)$ beyond the mean-field approximation with feedback of bosonic
fluctuations.  The minimum position of this potential, $\rho_0$, is related to
the superfluid density and acts as an order parameter for the
superfluid-to-normal phase transition. For convenience, and similar
to the relativistic case, we use the gap $\Delta_0^2 = g^2\rho_0$, rather than
$\rho_0$ itself as the order parameter. 

Note that the binding energy $\vare_{\rm B}<0$ is non-zero on the BEC side, and
the fermion chemical potential eventually becomes negative for large positive
scattering length. By contrast, we set $\vare_{\rm B}=0$ on the BCS side. The
quantity $\tilde{\mu}=\mu-\vare_{\rm B}/2>0$ is manifestly positive for non-vanishing
density, and we choose units such that $\tilde{\mu}=1$ when discussing
the whole crossover. 
A negative chemical potential shifts the minimum in the Sarma criterion
\eq{eq:Int2}. Taking this possibility into account, the criterion generalizes to
\begin{align}
 \label{eq:SarmaCrossover} \delta \mu > \min_{p} \sqrt{\vare_p^2+\Delta_0^2} =
  \begin{cases} \Delta_0\,, & (\mu\geq0)\\ \sqrt{\mu^2+\Delta_0^2}\,, &
(\mu<0)\end{cases}\,.
\end{align}

\subsection{Relation to the Relativistic Model}
\label{subsec:Comparison}

As we have argued above, the relativistic system \eq{eq:Lagrangianqmmuf} and the
non-relativistic one, \eq{eq:Lnonrel}, both describe two-component fermionic
systems in the BCS-BEC crossover. Actually, the Lagrangian
\eq{eq:Lagrangianqmmuf} can be seen as the relativistic analog of
\eq{eq:Lnonrel}: on a very basic level both represent a Yukawa system with
fermions coupled to two different chemical potentials. To be precise,
$\delta\mu$ in the non-relativistic case should be identified with the
quark chemical potential $\mu_q$, whereas the chemical potential $\mu$ in the
non-relativistic case corresponds to the isospin chemical potential $\mu_I$.
To simplify the comparison we provide a dictionary between the two systems in
\tab{tab:dictionary}.

On closer inspection, though, the field content of the models is
different: the relativistic spinor is subject to an additional chiral symmetry,
under which the left- and right-handed components,
$\psi_{R/L} = \frac12\left(1\pm\gamma_5\right)\psi$\,, transform separately.
The four (real) bosonic degrees of freedom, transforming as singlet and triplet
under isospin rotations, are related to these components in the following way
\begin{eqnarray*}
 \pi_+ & \sim & u_{L}d_{L}^\dagger + u_{R}d_{R}^\dagger\,,\\
 \pi_- & \sim & d_{L}u_{L}^\dagger + d_{R}u_{R}^\dagger\,,\\
 \pi_0 & \sim & u_{L}u_{L}^\dagger + u_{R}u_{R}^\dagger - (u\to d)\,,\\
 \sigma & \sim & u_{L}u_{R}^\dagger + u_{R}u_{L}^\dagger + (u\to d)\,, 
\end{eqnarray*}
where we have used the notation $ u := \psi_{u}$ and $ d := \psi_{d}$ for
better readability.
Using the correspondence $\psi_1 \leftrightarrow u$ and $\psi_2 \leftrightarrow 
d^\dagger$, it is clear that $\pi_+\leftrightarrow \phi$ and 
$\pi_-\leftrightarrow \phi^*\,.$
The other two bosonic degrees of freedom, $\pi_0$ and $\sigma$, however, have
no counterpart in the non-relativistic system. They reflect the larger symmetry
group, $SU(2)\times SU(2)$, of the relativistic system. Owing to this
discrepancy, one can expect that the impact of bosonic fluctuations on the
relativistic and non-relativistic systems is different. Furthermore, the
fields $u,\, d^\dagger$ each describe two independent fermions $u_L,
u_R$ and $d^\dagger_L, d^\dagger_R$, respectively, while $\psi_1$ and $\psi_2$
account only for a single fermion.

\begin{table}
  \begin{tabular}{c|c|l}
   non-relativistic & relativistic & interpretation \\\hline\hline
    $\psi_1,\, \psi_2$ & $\psi_u,\, \psi_d^\dagger$ & spin/flavor eigenstates
      \\ \hline
    $\mu$ & $\mu_I$ & induces pairing\\ \hline
    $\delta\mu$ & $\mu_q$ & knob to destroy pairing\\ \hline
 $\delta n =n_1-n_2$ & $\delta n = n_{q} - n_{\bar{q}}$ & population imbalance\\
    \hline
    $\Delta^2 = g^2\phi ^*\phi$ & $\Delta^2 = g^2 \pi_+\pi_-$ & pairing order
      parameter\\ \hline
    - & $\chi$ & chiral condensate 
  \end{tabular}
  \caption{Dictionary between quantities in the non-relativistic and
    relativistic system and their interpretation.}
  \label{tab:dictionary}
\end{table}

Other than that, the condensation of charged pions in \eq{eq:Lagrangianqmmuf} is
the equivalent of the di-fermion condensation occurring in the non-relativistic
system.
The related order parameter in both cases is the gap $\Delta$. 
Note that also the universal aspects of a second-order condensation transition
agree: the condensate $\Delta$ always breaks a $\mathrm{U}(1)$ symmetry.

\subsection{FRG Setting and Truncation}\label{sec:Trunc}

To investigate the stability of the Sarma phase in  the spin-imbalanced BCS-BEC
crossover we employ the FRG approach described in  \cite{Boettcher:2014a,
Braun:2014}.
We refer to those references for a detailed discussion of the truncation and
regularization scheme. Here we only summarize the main features which are
relevant for the present analysis.

To properly account for first-order phase transitions and the competition of
multiple minima one needs to know the effective potential
$U(\rho)$ over a large range of $\rho$-values. For this purpose we discretize
the potential on a grid in the gap $\Delta=g^2\rho$. Alternative approaches are
based on higher-order Taylor expansions of the effective potential around
a fixed value \cite{Pawlowski:2014zaa}, or the expansion in terms of suitable
basis functions. Note that a recent analysis of the BCS side in
\cite{Krippa:2014kra} is built on a Taylor expansion of $U(\rho)$ to order
$\rho^2$, as thus fails to resolve the first-order transition in the
perturbative Clogston limit \cite{PhysRevLett.9.266, Chandrasekhar}.
Furthermore, we want to remark that the FRG approach allows to
recover the mean-field result in a conceptually consistent way: neglecting the
bosonic contributions to the flow equations, the standard mean-field result is
reproduced \cite{PhysRevB.78.014522,PhysRevA.89.053630,Boettcher:2014a}.

From mean-field studies of the phase structure of the imbalanced UFG
\cite{PhysRevLett.96.060401,Gubbels:2006zz,2007NatPh...3..124P}, we expect a
first-order phase transition at low temperatures. This suggests that the
stability of the Sarma phase at zero temperature is decided according to
Scenarios II and III from above. To distinguish between these two scenarios,
renormalization effects on $\Delta_c$ and $\delta\mu$ are expected to be
important. In the present work, we introduce a single wave function
renormalization, $A_\phi$, for the bosonic field and set $A_\phi=Z_\phi$ in
\eq{eq:Lnonrel}. We do not include the particle-hole correction to
the four-fermion vertex or the renormalization of the fermion propagator. Those
contributions have been shown to be subleading for the phase structure of the
balanced system with the FRG \cite{PhysRevB.78.174528, PhysRevA.81.063619,
ANDP:ANDP201010458, PhysRevA.87.023606, PhysRevA.89.053630}.
Here we discuss why we expect them to be subleading for the discussion of
the existence of the Sarma phase as well.

From studies of the polaron \cite{PhysRevLett.97.200403, PhysRevLett.100.140407,
Schmidt:2011zu} and the balanced UFG \cite{PhysRevA.89.053630} it is known that
fluctuation effects tend to \emph{increase} the individual chemical potential,
$\mu_\sigma$, by a contribution approximately proportional to the chemical
potential of the other species, $\mu_{\bar{\sigma}}$. In both cases,
fluctuations induce renormalization effects on the order of $60\%$,
\begin{align}
\mu_{\sigma,\rm eff} \simeq \mu_\sigma+ 0.6\, \mu_{\bar{\sigma}}.
\end{align}
Assuming this relation to be generally valid, we can estimate the effective
imbalance to be given by
\begin{align}
  \delta\mu_{\rm eff} = (\mu_{1,\rm eff}-\mu_{2,\rm eff})/2 \simeq
  0.4\,\delta\mu\,,
\end{align}
i.e. the effective imbalance is smaller than the unrenormalized (or bare) one. The Sarma
criterion $\Delta_c < \delta\mu_{\rm eff}$, which has to be true
for the renormalized (or dressed) parameters, is even less likely fulfilled. In particular,
for most cases discussed below we find that the Sarma criterion is violated
already for the unrenormalized
imbalance. According to our argument here, this implies that it is also violated
for the renormalized one.

\begin{figure}
  \includegraphics[width=.5\textwidth]{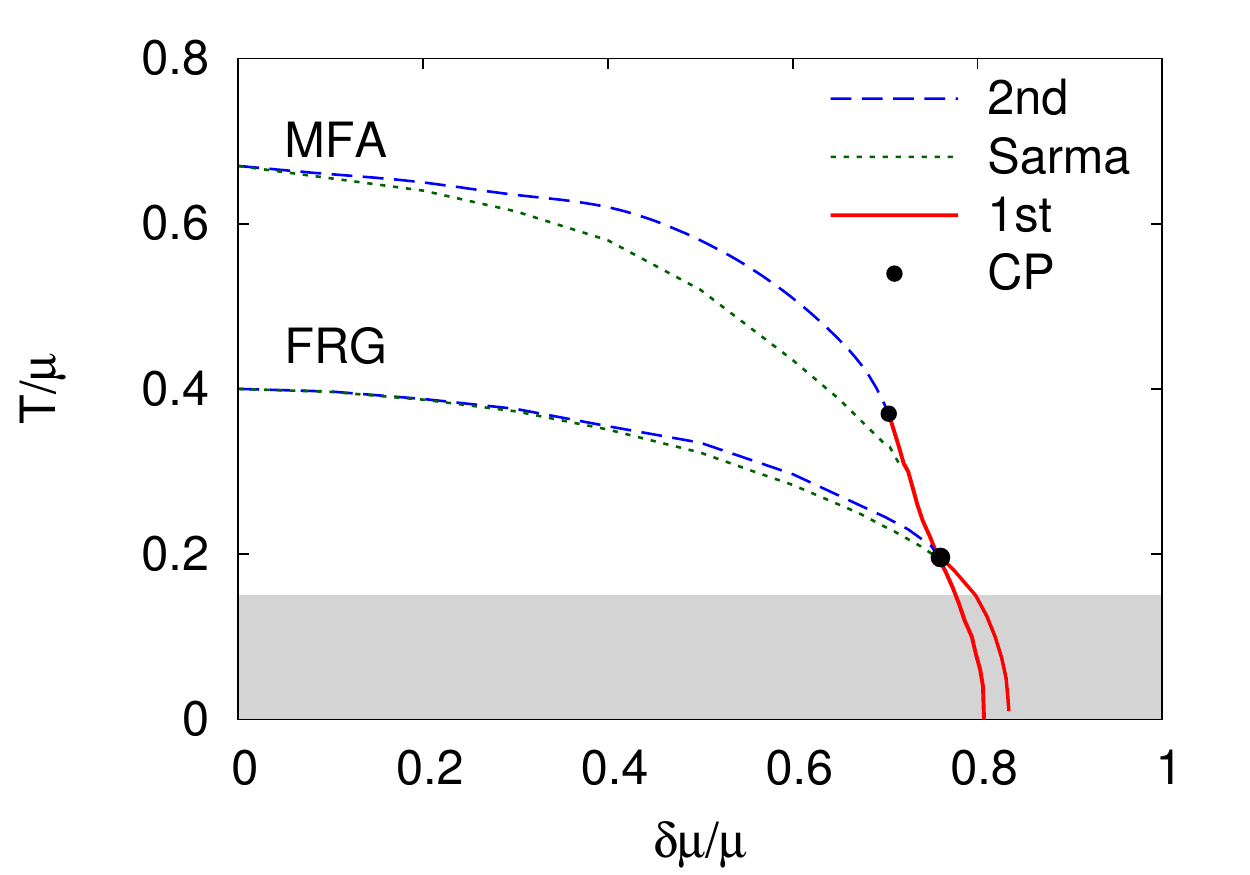}
  \caption{Phase structure of the imbalanced unitary Fermi gas. The upper
  lines correspond to the mean-field approximation (MFA), the lower ones to the
  FRG result. We observe a substantial decrease in the critical temperature due
  to bosonic fluctuations. 
  The Sarma condition $\Delta=\delta\mu$ is fulfilled along the dotted green
  line close to the second-order phase boundary. 
  Interestingly, in both cases we do not find a Sarma phase at zero
  temperature.}
  \label{fig:UFG_nonrel}
\end{figure}

\subsection{Unitary Fermi Gas}\label{sec:UFG}

Motivated by the similarity of its mean-field phase diagram to the relativistic
system discussed in \sec{sec:Relativistic}, we start our discussion with the
imbalanced UFG, where the superfluid is strongly correlated.

The mean-field phase structure is recovered by neglecting bosonic fluctuations
in the FRG flow equation. This is demonstrated in \fig{fig:UFG_nonrel} (upper
lines labelled ``MFA'').  We find a second-order phase transition (blue,
long-dashed line) for $\delta\mu=0$ in agreement with the expectation from the
balanced BCS-BEC crossover. The related mean-field critical temperature is
$T_c/\mu=0.665$. In the imbalanced case we observe a first-order transition
(red, solid line) from the superfluid to the normal phase at $\delta\mu_c/\mu =
0.807$ at zero temperature. The critical point, separating the first- from
the second-order transition line, is found at $(\delta\mu_{\rm CP}/\mu, T_{\rm
CP}/\mu) = (0.704, 0.373)$. These results are in line with the literature
\cite{PhysRevLett.96.060401,2007NatPh...3..124P}.
The Sarma phase (green, dotted line) appears in the vicinity of the second-order
transition line. Note that the dotted green line, corresponding to the condition
$\Delta=\delta\mu$, only serves as an orientation, since the transition is a
crossover at non-zero temperature. The Sarma crossover line terminates close
to the critical point, where it hits the first-order transition.
The jump in the gap then prevents the Sarma condition from being fulfilled for
lower temperatures. This corresponds to Scenario III discussed above. We
conclude that, at the mean-field level, there is no stable Sarma phase at $T=0$.

Next we include the feedback of bosonic fluctuations. 
The resulting phase diagram is also shown in \fig{fig:UFG_nonrel}
(lower lines labelled ``FRG''). At vanishing imbalance we again find a 
second-order phase transition. The transition temperature, however, is 
drastically reduced to $T_c/\mu=0.40\,.$  Overall, the inclusion of bosonic 
fluctuations makes the transition sharper, resulting in a {\it shrinking} Sarma 
phase. 
Furthermore, this phase appears at relatively high temperatures only. In this
regime the presence of gapless fermionic excitations is smeared out, and may be
difficult to detect in experiment.

At vanishing temperature we still find a first-order transition with a critical
imbalance of $\delta\mu_{\rm c}/\mu = 0.83$. This is larger than the mean-field
prediction, and in reasonably good agreement with the recent experimental
finding $\delta \mu_{\rm c}/\mu=0.89$ \cite{PhysRevA.88.063614}. The latter
reference also confirms the first-order phase transition at low temperatures.

Note that, due to its complexity, it is hard to evolve the FRG flow
for very small $k$ in the low temperature region. A conservative estimate for
the latter is indicated by the grey band in \fig{fig:UFG_nonrel}. The
determination of the phase boundary, however, is still reliable in this region.
A more detailed discussion of this point is provided in \cite{Boettcher:2014a}.
Furthermore, the critical point, the onset of the first-order transition and the
end of the Sarma phase all lie well above this band. Hence we can draw our
conclusions independent of this limitation.

\begin{figure}
  \includegraphics[width=.5\textwidth]{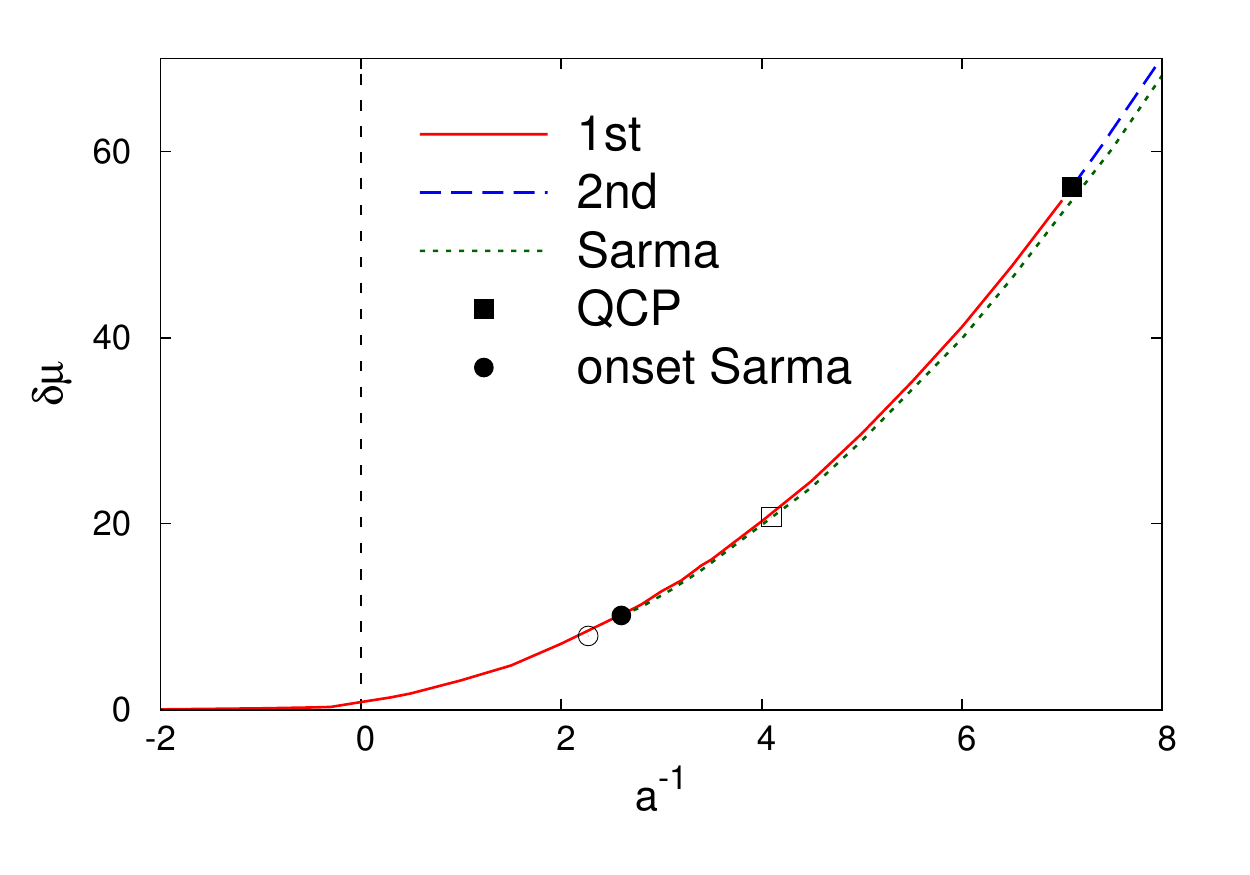}
  \caption{Quantum phase diagram of the imbalanced BCS-BEC crossover from 
    the FRG. Units are chosen such that $\tilde{\mu}=\mu - \epsilon_B/2 = 1\,.$ The
    first-order superfluid phase transition appearing on the BCS side persists
    on the BEC side up to the quantum critical point (QCP).
    The QCP is marked by a filled (open) square for the result from the FRG
    (mean-field) analysis. The onset of the Sarma phase along the first-order
    line according to Scenario II is indicated by the filled (open) circle for
    the FRG (mean-field) result. The boundary of the Sarma phase on the BEC side
    is given by the dotted green line.}
  \label{fig:T0_phase}
\end{figure}

\subsection{BCS-BEC crossover}\label{sec:Cross}

Our initial motivation to study the fate of the Sarma phase at
low temperatures upon inclusion of fluctuations was the claim that the
relativistic system discussed in \sec{sec:Relativistic} agrees with the UFG on
the mean-field level. However, the mean-field phase diagrams throughout the
BCS-BEC crossover look very similar, apart from a change of scales.
Moreover, the simple estimate presented in \sec{sec:Relativistic}
suggests that the relativistic phase diagram shown there rather corresponds
to a point on the BCS side of the crossover. Hence we
extend our study to finite scattering lengths (the only tunable parameter
in this system) in order to identify a region that might support a stable Sarma
phase at $T=0$\,. Additionally we want to note that even the phase structure of
the imbalanced BCS-BEC crossover beyond mean-field had been unknown to a large
extent so far.

In the following we focus on the phase structure of the
imbalanced BCS-BEC crossover at zero temperature, i.e. the quantum phase
diagram. The mean-field result has previously been calculated in, e.g., 
\cite{PhysRevLett.96.060401,2007NatPh...3..124P, PhysRevA.74.033606}. The
superfluid-to-normal transition is of first order on the BCS side ($a^{-1}<0$).
This behavior persists on the BEC side ($a^{-1}>0$) up to a quantum critical
point (QCP) where the transition turns to second order. Within the mean-field
approximation, the QCP is located at 
$$(\sqrt{\tilde{\mu}}a)^{-1}_{\rm MF}=4.19\,,\quad \delta\mu_{\rm  MF}=21.6\tilde{\mu}=0.61|\vare_{\rm
B}|\,.$$ 
We again employ the definition $\tilde{\mu}=\mu-\vare_{\rm B}/2$.

The quantum phase diagram including the feedback of bosonic fluctuations from
Functional Renormalization is shown in \fig{fig:T0_phase}. 
Its structure is even quantitatively very similar to the mean-field result,
hence we only show the FRG result and have superimposed the locations of the QCP
and the onset of the Sarma transition from the MFA. On the BCS side and in the
vicinity of the resonance, the transition is of first order. On the BEC side
there is a QCP where a second-order line emerges. Its
coordinates read 
$$(\sqrt{\tilde{\mu}}a)^{-1}_{\rm FRG}=7.1\,,\quad \delta\mu_{\rm FRG}=56.2\tilde{\mu}=0.56|\vare_{\rm
B}|\,,$$ 
within our approximation. We see that, contrary to the relativistic case
discussed above, fluctuations rather induce a first-order phase transition than
a second order one.

The onset of the Sarma phase on the BEC side is located to the left of the QCP,
and thus happens according to Scenario II in the terminology introduced above.
The boundary of the Sarma phase according to \eq{eq:SarmaCrossover} is
indicated by the dotted green line in \fig{fig:T0_phase}. It terminates in
the first-order line at $(\sqrt{\tilde{\mu}}a)^{-1}_{\rm MF}=2.27$ in MFA, which is shifted towards
$(\sqrt{\tilde{\mu}}a)^{-1}_{\rm FRG}=2.6$ when including bosonic fluctuations. To the right of the QCP we
always find a stable Sarma phase below the second-order line, according to
Scenario I. Since the Sarma phase only appears on the BEC side, the
corresponding magnetized superfluid constitutes a homogeneous state consisting
of a BEC of diatomic molecules with excess majority atoms.

Hence we find that our initial question for a parameter set that
supports a Sarma phase at $T=0$ including fluctuations, but not on the
mean-field level, has to be answered negatively: the onset of the Sarma phase
occurs at lower inverse scattering lengths in the MFA than with the FRG. In
contrast, there is an, albeit small, window where the mean-field phase diagram
shows a Sarma phase which vanishes after the inclusion of fluctuations.
Moreover, as distinguished from the relativistic case, a Sarma
phase arises only on the \emph{BEC side} of the crossover in the
non-relativistic system.

\section{Conclusions}\label{sec:Concl}

Triggered by the close resemblance of the phase diagrams of relativistic and
non-relativistic two-component fermion systems on the mean-field level, we have
studied the fate of this similarity once fluctuations are taken into account.

We have studied the two-flavor quark-meson model coupled to quark as
well as isospin chemical potentials as a relativistic realization of this
system. Fixing the isospin chemical potential to a
value that allows for pion condensation, $\mu_I>m_\pi/2$, one can study the
breakdown of the related pion-superfluidity as the quark chemical potential is
varied. In fact, on the mean-field level the resulting phase diagram looks
remarkably similar to the one of the spin-imbalanced unitary Fermi gas, see
\sec{sec:NonRelativistic}. Including fluctuations in the relativistic setting,
the phase diagram changes drastically for $\mu_I\gtrsim 0.79m_\pi$: At low
temperatures, the transition line splits into two branches, one of first and one
of second order. Interestingly, the Sarma-crossover line meets the first-order
transition in the critical point, while the second-order line continues down to
$T=0$. This means that the relativistic system features a Sarma phase at $T=0$.

The non-relativistic system under consideration, the spin-imbalanced BCS-BEC 
crossover of ultracold atoms, does not show this feature despite the apparent
similarities in the mean-field phase structure of both system. While the
location of the phase boundaries is modified by the inclusion of fluctuations,
e.g. the critical temperature is drastically reduced, the general structure of
the phase diagram persist. At unitarity, the unpolarized superfluid
ground state at zero temperature is separated from the normal phase by a 
first-order phase transition. Moreover, the Sarma criterion (\ref{eq:Int2}) is 
not fulfilled at the phase boundary. Thus the superfluid with unequal densities
$n_1\neq n_2$ can only be realized as an inhomogeneous mixed state. The
realization of a first-order phase transition is in agreement with
the available experimental data. 

Since the relativistic system considered in \sec{sec:Relativistic} presumably
lies on the BCS side of the crossover, we have extended our non-relativistic
study to the whole BCS-BEC crossover at low $T$. 
A zero temperature Sarma phase in the non-relativistic setup  is only found on
the  BEC side of the crossover, close to the region predicted from mean-field
theory. We were able to locate the QCP on the BEC side, where the transition
changes from first to second order.  
An interesting question then concerns the critical exponents at the QCP on the
BEC side, which can be computed with the FRG to high accuracy
\cite{Litim:2002cf}.
A more detailed analysis of the quantum critical properties of the QCP will be
presented elsewhere.

Finally, as we have discussed in \sec{subsec:Comparison}, there exist some
crucial differences between the relativistic and non-relativistic
systems under consideration: Due to the additional chiral symmetry, the
relativistic system features effectively four species of fermions as well as
four real bosonic degrees of freedom.
Of these, the chirality-preserving but flavor-mixing combinations, corresponding
to the bosonic fields $\pi_{\pm}$, are the counterparts of the complex
non-relativistic boson $\phi$\,. The presence of two additional bosonic modes,
however, modifies the dynamics of the system substantially. In particular, it
results in a Sarma phase at vanishing temperature in the relativistic theory
that is not present in the non-relativistic setting. The agreement of the phase
diagrams on the mean-field level, on the other hand, suggests that the
differences in the fermionic sector are not as crucial.

Based on these observations, we can now suggest a non-relativistic system that
resembles the relativistic one more closely. Such a system would be interesting
to study both experimentally and theoretically, since it might feature a Sarma
phase at $T=0$, similar to the relativistic theory discussed above. For this
purpose, one needs to study a system with four fermion species,
$\psi_{1,2,3,4}$. Furthermore, interactions need to be tuned such that channels
equivalent to the interactions in \eq{eq:Lagrangianqmmuf} are present  and have
similar interaction strengths. The corresponding microscopic Hamiltonian, which
needs to be implemented with cold atoms, reads
\begin{align}
  \hat H =  \int \mbox{d}^3 x \Biggl[& \sum_{\sigma=1}^4 \psi_\sigma^\dagger
\Bigl(-\frac{\nabla^2}{2M}-\mu_\sigma\Bigr)\psi_\sigma\\
 \nonumber &\mbox{ }+\lambda\Bigl((\psi_1\psi_2)^\dagger\psi_1\psi_2 +
(\psi_3\psi_2)^\dagger\psi_3\psi_2 \\
 \nonumber & \mbox{ }  + (\psi_1\psi_4)^\dagger\psi_1\psi_4 +
    (\psi_3\psi_4)^\dagger\psi_3\psi_4\Bigr)\Biggr]\,.
\end{align}
The resulting system then possesses a chiral symmetry similar to the
relativistic one, with $SU(2)_L$ acting on the doublet $(\psi_1,\psi_3)$ and
$SU(2)_R$ acting on $(\psi_2,\psi_4)$. The interaction involves particular
combinations of $n_\alpha~=~\psi_\alpha^\dagger\psi_\alpha$, namely $(n_1 +
n_3)(n_2+n_4)\,.$ One could expect to find a similar phase structure as shown
in \fig{fig:sarmarel}, in particular a Sarma phase at $T=0$.

\section*{Acknowledgements}
The authors thank J. Braun and D. Roscher for valuable discussions
and collaboration on related work. I. B. acknowledges funding from
the Graduate Academy Heidelberg. L.v.S. is supported by the Helmholtz
International Center for FAIR  within the LOEWE program of the State of
Hesse. This work is also supported by the Helmholtz Alliance HA216/EMMI and the
grant ERC-AdG-290623.


\bibliographystyle{apsrev4-1}
\bibliography{references_sarma} 

\end{document}